# Tracing the main elements and electron orbitals that induce superconducting phase transition

Sheng-Hai Zhu [a], Han Qin [b], Mi Zhong [a], Dai-He Fan [a], Xiang-Hui Chang [a], Yun Wei [a], Miao Zhang [a], Tao Zhu [c], Bin Tang [d], Fu-Sheng Liu [a], Qi-Jun Liu [a] [*]

[a] School of Physical Science and Technology, Southwest Jiaotong University, Key Laboratory of Advanced Technologies of Materials, Ministry of Education of China, Chengdu 610031, China

[b] School of Science, Xihua University, Chengdu 610039, People's Republic of China

[c] State Key Laboratory of Traction Power, Southwest Jiaotong University, Chengdu 610031, People's Republic of China

[d] State Key Laboratory of Solidification Processing, Northwestern Polytechnical University, Xi'an 710072, China

Correspondence about the paper should be at the following address and e-mail address:

School of Physical Science and Technology, Southwest Jiaotong University, Chengdu, Sichuan 610031, China

Qi-Jun Liu, E-mail: qijunliu@home.swjtu.edu.cn

---

[*] **Corresponding author. qijunliu@home.swjtu.edu.cn**

**ABSTRACT**

The experimental determination of the superconducting transition requires the observation of the emergence of zero-resistance and perfect diamagnetism state. Based on the close relationship between superconducting transition temperature ($T_c$) and electron density of states (DOS), we take two typical superconducting materials Hg and $ZrTe_3$ as samples and calculate their DOS versus temperature under different pressures by using the first-principle molecular dynamics simulations. According to the analysis of the calculation results, the main contributors that induce superconducting transitions are deduced by tracing the variation of partial density of states near $T_c$. In particular, the microscopic mechanism of pressure increasing $T_c$ is further analyzed.

The transition to a zero-resistance and perfect diamagnetism state is the macroscopic manifestation of superconductivity. With the introduction of Bardeen-Cooper-Schrieffer (BCS) theory, the microscopic mechanism of superconducting phenomenon has been preliminarily explained [1]. BCS mentioned that the electron density of states (DOS) is one of the important factors affecting the superconducting transition temperature ($T_c$). Because of the important influence of electro-phonon coupling, researchers have repeatedly studied the relationship between the DOS and superconductivity [2, 3]. Although the discovery of more and more unconventional superconducting materials has led researchers to realize that the BCS theory is not applicable to all superconductors [4-7], the close relationship between the electron DOS and the superconducting phenomenon is still undeniable [8-12]. In these excellent studies, the DOS variation of the electron orbitals for each atom in superconductors near the $T_c$ is rarely mentioned, but it is of great significance for exploring the microscopic mechanism of superconducting transition.

Due to the inseparable connection between the electron DOS and superconductivity, we believe that the DOS of superconducting materials above and below $T_c$ will definitely be significantly different. In order to find out this difference here, two representative samples (Hg and ZrTe$_3$) are selected for our present research and we look forward to further analyzing the main factors that induce superconducting transitions in superconductors on this microscopic basis. The Hg is a substance that can never be bypassed when referring to superconductivity, Onnes first discovered the zero-resistance state of Hg at an extremely low temperature of 4.2 K in 1911, which opened the era of superconductivity. If Hg is the source of the superconducting phenomenon, then ZrTe₃ is one of the superconducting materials currently standing on the front line. ZrTe₃ has attracted extensive interest due to the coexistence and competition between charge density waves and superconducting states under high pressures [13-15], researchers have employed experimental methods and theoretical calculations to investigate it in anticipation of learning more about the truth of superconductivity [16-18]. It is worth noting that the $T_c$ of ZrTe3 increases with the increasing pressure [13, 16], and the exploration of DOS for ZrTe3 may provide effective help to explain the micro mechanism of pressure promoting $T_c$.

In view of the need to obtain the physical properties of Hg and ZrTe₃ at specified pressures and temperatures, molecular dynamics (MD) simulation is considered to be an effective approach to advance the research [19, 20]. In this paper, the electron DOS of these two materials under different pressures and temperatures are investigated using the first-principle MD simulations.

The CASTEP code [21] based on the density-functional theory framework with the GGA-PBE functional [22] was employed to perform molecular dynamics simulations adopting isobaric-isothermal (*NPT*) ensemble. The total simulation time for each temperature point was 5ps, with time steps of 0.5 fs. The electron–core interactions were solved using ultrasoft and norm-conserving pseudopotentials for Hg and ZrTe₃, respectively. The Hg $5s^25p^65d^{10}6s^2$, Zr $4s^24p^64d^25s^2$, and Te $5s^25p^4$ electrons were described as valence electrons.

According to the superconducting transition temperature of 4.2 K, the electron

DOS of Hg near the $T_c$ is calculated using MD simulations under 0 GPa. The differences in DOS at the micro temperature interval scale are so small that they almost overlap in most areas. Therefore, the 3D waterfall diagram is adopted to more clearly show the DOS curves of Hg corresponding to different temperatures, as shown in Figure 1. It is not difficult to observe that as the temperature gradually decreases, the curve hardly changes in the initial period of cooling. When the temperature continues to drop to 4K, the total DOS curve shows a significant change. The main fluctuations are concentrated in the zone from −8 eV to −5 eV, as the temperature changes from 5K to 4K, the DOS dispersion increases while the peaks of DOS decrease in this region. This phenomenon indicates that the metallicity of Hg is enhanced and the electrons become more active. As the temperature continues to decrease from 4K, the total DOS curve no longer shows visible variation. Corresponding to the $T_c$ of 4.2K measured in the experiment, it is reasonable to assume that the variation of DOS curve here is closely related to the superconducting transition. The total DOS of the crystal is composed of the partial density of states (PDOS) of the different orbitals for all atoms, so the change of the total DOS should be the overall manifestation of the variation in the PDOS.

In order to deduce the main contributors that induce superconducting transition, the crystal structure of Hg and the calculated PDOS of the two temperature points near $T_c$ are also shown in Figure 1. Trigonal Hg belongs to space group $R\bar{3}m$, all mercury atoms are in the same position. With regard to the PDOS curves of different orbitals, the region from −8 eV to −5 eV is dominated by the d orbital. It can also be observed from the figure that the change of the total DOS near $T_c$ is mainly caused by the variation of the d orbital. Based on the above discussion, we conclude that the d orbital is the main contributor to the superconducting transition of Hg. Admittedly, since the two outermost electrons of Hg are in the d orbital, the inference here is actually predictable. Going a step further, by replacing the sample with a polyatomic molecule, the method of tracking the DOS will play a more creative role.

Unlike mercury, its transition to superconducting state at atmospheric pressure, ZrTe$_3$ behaves superconducting only at high pressure [13, 23]. The $T_c$ of ZrTe$_3$ measured by Gu et al. in the experiment [16] is 4.3 K, 6 K and 7.1 K under the pressure of 8.2 GPa, 15 GPa and 27.7 GPa, respectively. Meanwhile, they concluded that the $T_c$ of ZrTe$_3$ increases with increasing pressure and reaches a maximum at 27.7 GPa. To further confirm our idea, we calculate the DOS of ZrTe$_3$ versus temperature under pressures of 10 GPa and 20 GPa, as shown in Figure 2. At the initial stage of cooling, the total DOS under these two different pressures hardly changes as the temperature decreases. When the temperature continues to drop, the DOS under 10 GPa and 20 GPa changes during the process of 5 K to 4 K and 7 K to 6 K, respectively. After the critical temperature is crossed, the further decreasing temperature loses its influence on the DOS curves. The results are highly consistent with the $T_c$ given by the experiment [16]. With regard to the variation of total DOS under 10 GPa, the peak in the zone of −2 eV to −1 eV shows a visible drop when the temperature drops from 5 K to 4 K, while no obvious changes are observed in other regions. Focus on the total DOS curves under 20 GPa, in addition to the change in the region from −2 eV to −1 eV, the peak in the zone of 1 eV to 2 eV sharpens when the temperature is reduced from 7 K to 6 K.

The calculated PDOS of these regions where the total DOS has changed is shown in Figure 3. Further analysis of PDOS curves under 10 GPa, it can be observed from Figure 3a that as the temperature drops from 5K to 4K, the PDOS of each atom shows visible changes in the region from −2 eV to −1 eV. Among them, the variations of Te3 are the most obvious, and the splitting phenomenon appears in both Te3-s and Te3-p orbitals. In contrast, the change in Te2 is relatively small compared to the atoms in the other three positions. For the PDOS from 7 K to 6 K at 20 GPa, Figure 3b shows that the curves of Te1 and Te3 in the region from −2 eV to −1 eV change more obviously. As shown in Figure 3c, Zr-p/d, Te1-s, Te2-s/p and Te3-s all contribute to the sharpening of peak near 1.7 eV.

Combined with the above discussion and the crystal structure of $ZrTe_3$ shown in Figure 2, we deduce that the drop in temperature under the pressure of 10 GPa mainly affects the atoms near the crystal edge, especially Te3, which ultimately leads to the transition of $ZrTe_3$ to the superconducting state. As the pressure rises to 20 GPa, the sensitivity of the material increases, and Te2 in the middle of the crystal can be affected by both temperature and pressure, so that the entire structure can be changed under the influence of external conditions. This leads to the early arrival of the superconducting transition, which is manifested as an increase in $T_c$ under higher pressure. We believe that this inference can explain the microscopic mechanism of pressure increasing the $T_c$ in most superconducting materials. So how does the temperature under different pressures affect the atoms at various positions to change their respective PDOS?

The atomic motion path of ZrTe$_3$ near $T_c$ is shown in Figure 4. For the path from 5 K to 4 K at 10 GPa, the movement of Te3 is the most obvious, which is consistent with the above discussion on PDOS. Both Te2 and Zr move slightly, and they almost have the same movement track. Since DOS reflects the bonding characteristics between atoms, this result for Zr does not correspond directly to PDOS analysis. As shown in Figure 2, Te2 only bonds with Zr, that is, when the two atoms have the same moving trajectory, it has little effect on the Zr-Te2 bond, so the PDOS of Te2 discussed above hardly changes. For Zr, it simultaneously bonds with Te1, Te2, and Te3. Although the distance of its own movement is slight, the PDOS of Zr will also show significant changes as Te1 and Te3 move obviously. With regard to the atomic motion path of ZrTe$_3$ under 20 GPa, all atoms except Zr have a visible movement, which is in good agreement with the analysis of PDOS. Therefore, tracking the variation of PDOS in polyatomic molecule plays a more creative role. This method can effectively deduce the main contributors to the superconducting transition by exploring the causes of the total DOS change near $T_c$, and make a further inference on the superconducting transition mechanism at the micro level.

In the final analysis, temperature and pressure are only used as external conditions to affect the properties of materials. We have a bold conjecture that the superconducting state is only directly related to the structure of the material itself. Unfortunately, the superconducting structures required by all potential superconductors do not exist in a natural state under normal temperature and pressure [4, 24]. Therefore, the superconducting properties are basically exhibited under extreme pressure and temperature conditions. The harsh external conditions also hinder the wider application of superconductors. By imposing other external intervention measures to replace the effect of extreme pressure and temperature on the structure, so as to realize the transformation of the superconducting state within an acceptable temperature and pressure range. We think this will be a path worthy of further exploration.

In summary, we have investigated the DOS of Hg and $ZrTe_3$ versus temperature under different pressures by employing molecular dynamics simulations. All of their DOS show visible changes around their respective $T_c$, while the DOS curves in other temperature ranges are hardly affected by the temperature difference at small intervals. By tracing PDOS, the factors that caused the change of total DOS are found, thus determining the main contributors that induce the transition of these two samples to the superconducting state. Furthermore, the superconducting mechanism that the higher pressure increases $T_c$ of $ZrTe_3$ is analyzed. The pressure increases the sensitivity of materials, so that the change in temperature has a more significant impact on the structure, which leads to the early arrival of the $T_c$. Based on the above results and discussions, we propose a conjecture that needs to be further demonstrated, that is, the superconducting state is only directly related to the material structure.

**AUTHOR STATEMENT**

Sheng-Hai Zhu performed the data analysis and wrote the paper; Han Qin and Mi Zhong contributed to investigations, methodology, writing - review & editing; Dai-He Fan, Xiang-Hui Chang, Yun Wei, Miao Zhang and Tao Zhu contributed to the theoretical analysis, writing - review & editing; Bin Tang and Fu-Sheng Liu contributed to investigations, methodology and software; Qi-Jun Liu designed the study, conceptualization, project administration, resources, supervision. All authors contributed to the general discussion.

**ACKNOWLEDGMENTS**

This work was supported by the National Natural Science Foundation of China (Grant No. 51402244), the Fundamental Research Funds for the Central Universities (Grant Nos. 2682019LK07).

**DATA AVAILABILITY**

The data that support the findings of this study are available from the corresponding author upon reasonable request.

**Figure 1.** The calculated DOS and PDOS of Hg versus temperature under 0 GPa and the crystal structure of Hg.

**Figure 2.** The calculated DOS of ZrTe$_3$ versus temperature under different pressures and the crystal structure of ZrTe$_3$.

**Figure 3.** The calculated PDOS of ZrTe$_3$ near $T_c$ under pressures of (a)10 GPa and (b) (c)20 GPa.

**Figure 4.** The atomic motion path of ZrTe$_3$ in 3D image near $T_c$ under pressures of 10 GPa and 20 GPa.

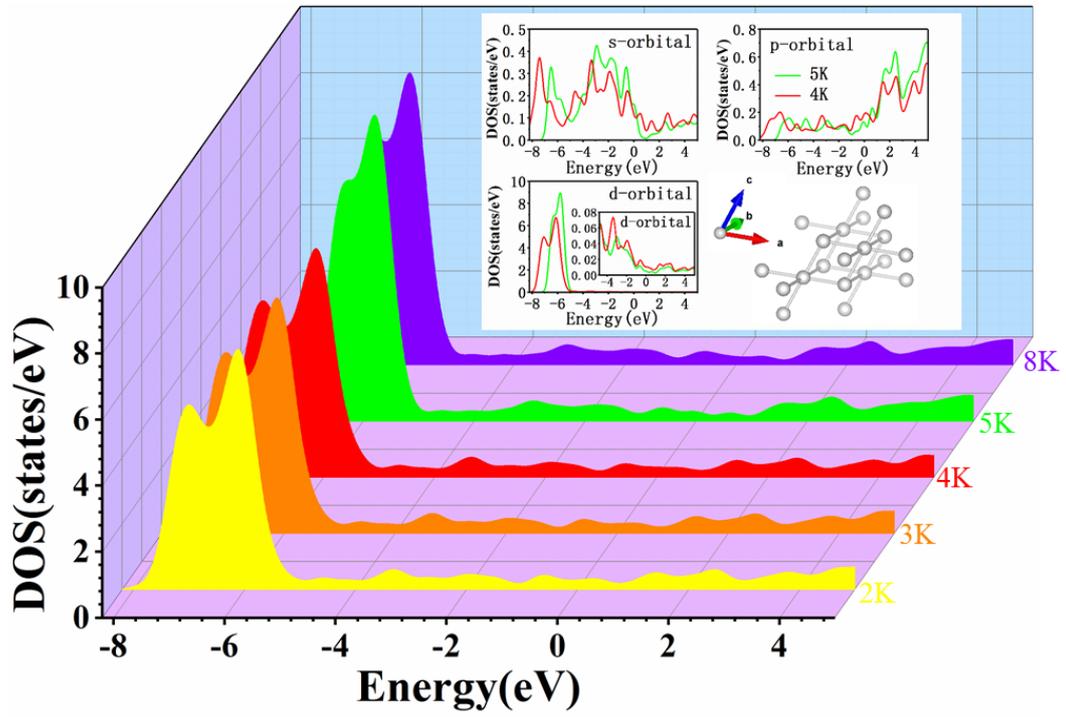

**Figure 1.** The calculated DOS and PDOS of Hg versus temperature under 0 GPa and the crystal structure of Hg.

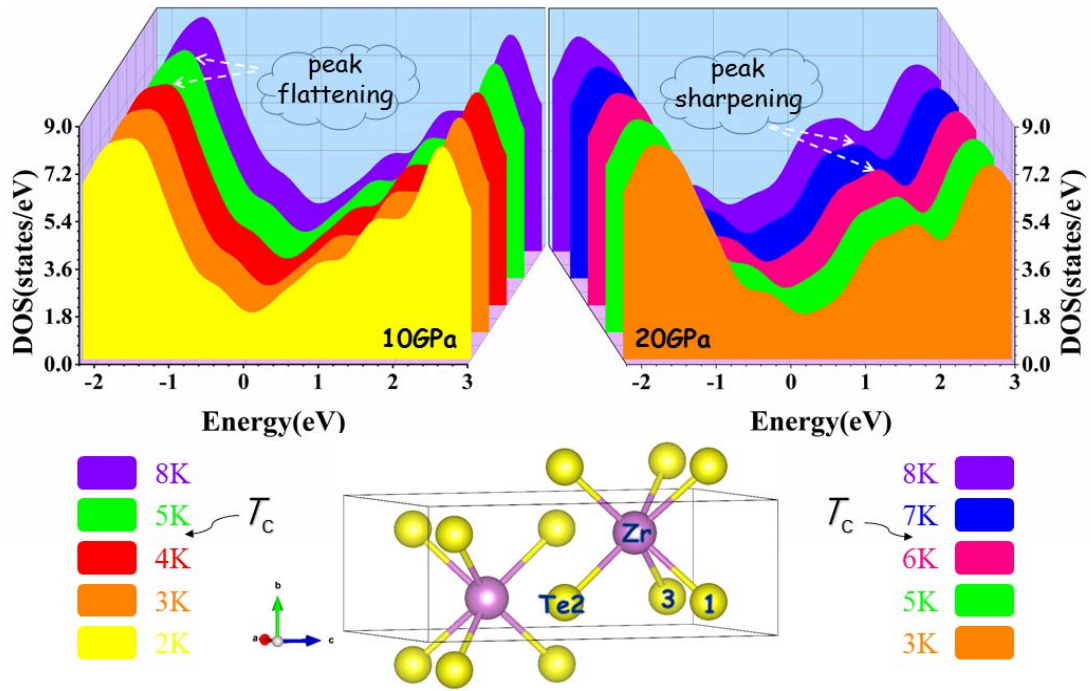

**Figure 2.** The calculated DOS of ZrTe$_3$ versus temperature under different pressures and the crystal structure of ZrTe$_3$.

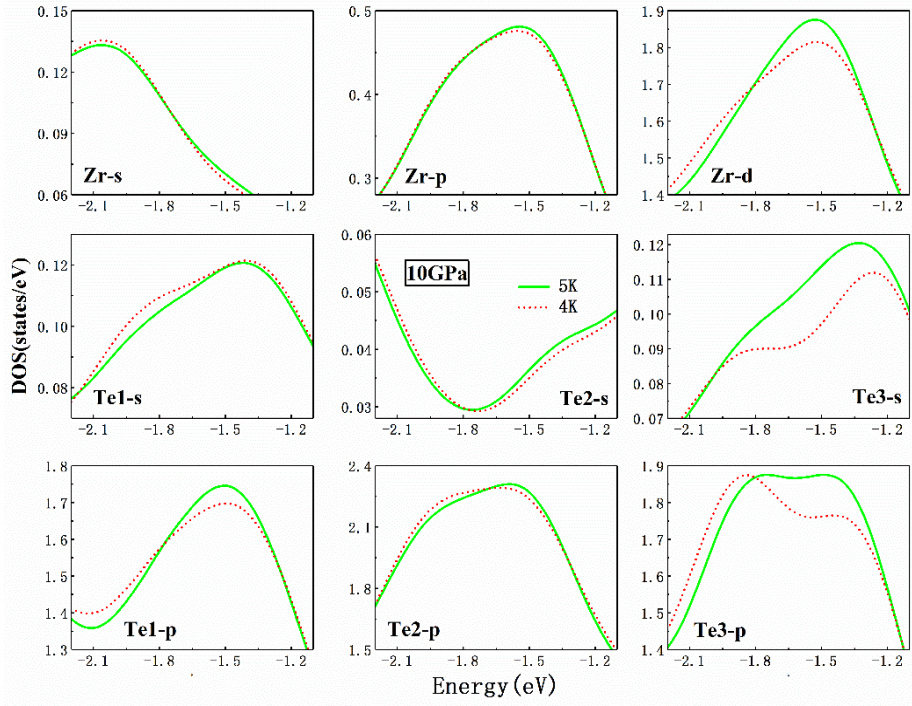

(a)

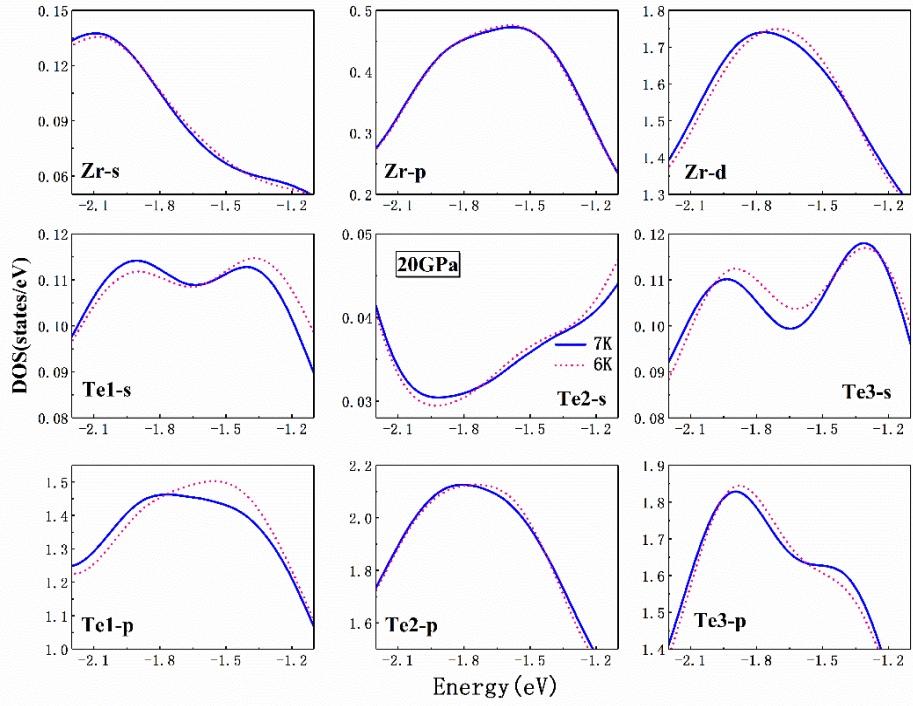

(b)

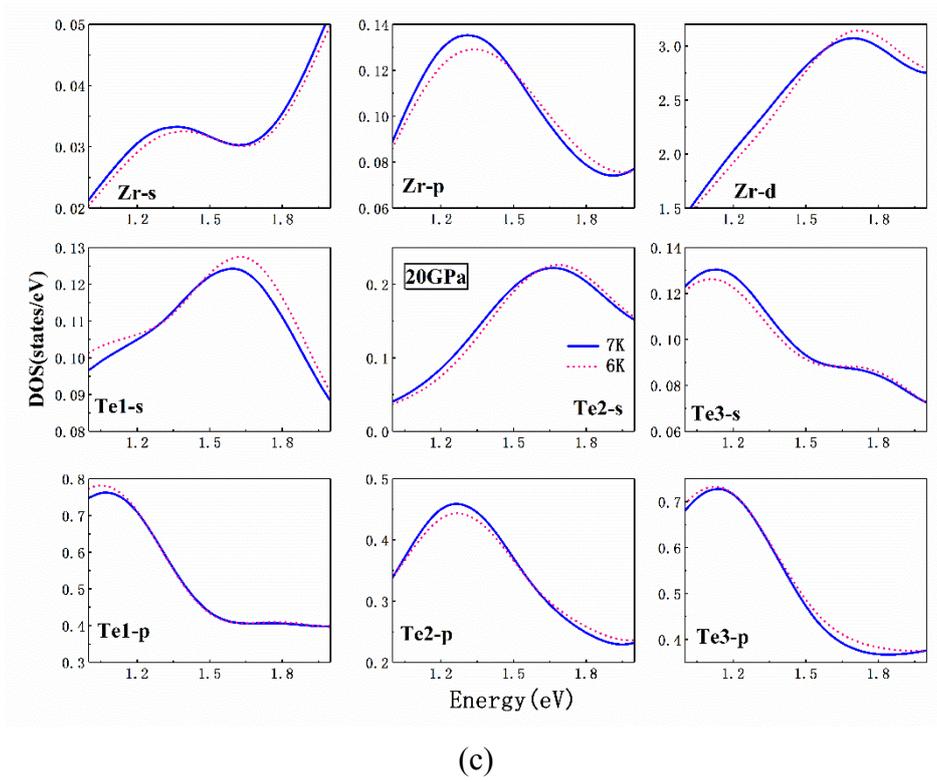

(c)

**Figure 3.** The calculated PDOS of ZrTe$_3$ near $T_c$ under pressures of (a)10 GPa and (b) (c)20 GPa.

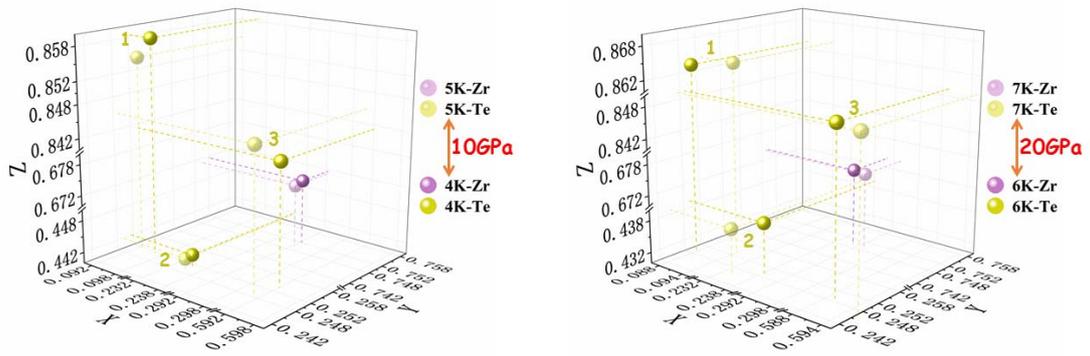

**Figure 4.** The atomic motion path of ZrTe$_3$ in 3D image near $T_c$ under pressures of 10 GPa and 20 GPa.